\documentclass[twocolumn,aps,prd,nofootinbib]{revtex4-1}
\usepackage[colorlinks=true,linkcolor=black,citecolor=blue,urlcolor=black]{hyperref}
\usepackage{graphicx,soul}
\usepackage{subfigure}
\usepackage{natbib}
\usepackage{epsfig}

\usepackage{amstext,amsmath,amsfonts,amssymb,amsthm}
\usepackage[format=plain,justification=centerlast]{caption}
\usepackage{color}

\usepackage{verbatim}
\usepackage{lipsum}
\usepackage{float}
\usepackage{enumitem}
\usepackage{color,array,wrapfig}
\usepackage{caption}
\def\be{\begin{equation}}
\def\ee{\end{equation}}
\def\bes{\begin{eqnarray}}
\def\ees{\end{eqnarray}}
\def\ba{\begin{align}}
\def\ea{\end{align}}
\def\bwt{\begin{widetext}}
\def\ewt{\end{widetext}}

\newbox\one
\newbox\two
\long\def\loremlines#1{%
    \setbox\one=\vbox {%
       Test.\footnote{a footnote}%
      \lipsum\footnote{Another footnote.}%
     }
   \setbox\two=\vsplit\one to #1\baselineskip
   \unvbox\two}

\begin{document}
\title{Constraining Cosmological and Galaxy Parameters using \\Strong Gravitational Lensing Systems.}

\author{Darshan Kumar$^{1}$}
\email{dkumar1@physics.du.ac.in, darshanbeniwal11@gmail.com}
\author{Deepak Jain$^{2}$} 
\email{djain@ddu.du.ac.in} 
\author{Shobhit Mahajan$^{1}$}
\email{sm@physics.du.ac.in}
\author{Amitabha Mukherjee$^{1}$} 
\email{ amimukh@gmail.com} 
\author{Nisha Rani$^{3}$} 
\email{ nisharani3105@gmail.com} 
\affiliation{$^{1}$Department of Physics \& Astrophysics, University of Delhi, Delhi$-$110007 India.}
\affiliation{$^{2}$ Deen Dayal Upadhyaya College, University of Delhi, Dwarka, New Delhi$-$110078 India.}
\affiliation{$^{3}$ Miranda House, University of Delhi, University Enclave, Delhi$-$110007 India.}

\vskip 1cm
\begin{abstract}
Strong gravitational lensing along with the distance sum rule method can constrain both cosmological parameters as well as density profiles of galaxies without assuming any fiducial cosmological model. To constrain galaxy parameters and cosmic curvature $(\Omega_{k0})$, we use the distance ratio data from a recently compiled database of  $161$ galactic scale strong lensing systems. We use databases of supernovae type-Ia (Pantheon) and Gamma Ray Bursts (GRBs) for calculating the luminosity distance. To study the model of the lens galaxy, we consider a general lens model namely, the Extended Power-Law model. Further, we take into account two different parametrisations of the mass density power-law index $(\gamma)$ to study the dependence of $\gamma$ on redshift. The best value of $\Omega_{k0}$ suggests a closed universe, though a flat universe is accommodated at $68\%$ confidence level. We find that parametrisations of $\gamma$ have a negligible impact on the best fit value of the  cosmic curvature parameter.

Furthermore, measurement of time delay can be a promising cosmographic probe via the ``time delay distance" that includes the ratio of distances between the observer, the lens and the source. We again use the distance sum rule method with time-delay distance dataset of H0LiCOW to put constraints on the Cosmic Distance Duality Relation (CDDR) and the cosmic curvature parameter $(\Omega_{k0})$. For this we consider two different redshift-dependent parametrisations of the distance duality parameter $(\eta)$. The best fit value of $\Omega_{k0}$ clearly indicates an open universe. However, a flat universe can be accommodated at $95\%$ confidence level. Further, at $95\%$ confidence level, no violation of CDDR is observed. We believe that 
a larger sample of strong gravitational lensing systems is needed in order to  improve the  constraints on the cosmic curvature and distance duality parameter.    

\end{abstract}

\maketitle
\section{Introduction}
A fundamental problem in modern cosmology is to determine whether the universe is spatially open, flat or closed. This is because the curvature of the universe plays an important role in its evolution. The most stringent constraint on the cosmological curvature parameter comes from the latest Planck result (2018) under the assumption of the $\Lambda$CDM model. The result supports a spatially flat universe with a high confidence level \cite{planck2018}. About a decade ago, Clarkson et al. (2008) proposed a model-independent method to measure the cosmological curvature parameter $(\Omega_{k})$ \cite{cc2008}. But the problem is that it involves derivatives of the distance w.r.t. the redshift which introduces a considerable amount of uncertainty in the estimated curvature parameter. Recently, R$\ddot{\textnormal{a}}\textnormal{s}\ddot{\textnormal{a}}$nen et al. proposed a model-independent method called the distance sum rule which is based on the assumption of the validity of the FLRW metric \cite{sr2015}. Any violation in the distance sum rule directly hints at a violation of the FLRW metric. The distance sum rule could be helpful to put constraints on the curvature parameter if it is found consistent  with various observational datasets.

In recent years, Strong Gravitational Lensing (SGL) has become a powerful technique to test various assumptions and relations in cosmology \cite{dw1979}. One can study both cosmological and galaxy parameters using lens systems. Observations of SGL can provide information of a distance ratio $d_A^{^{ls}}/d_A^{^{os}}$, where $d_A^{^{ls}}$ and $d_A^{^{os}}$ are the angular diameter distances between the source-lens and the observer-source respectively. To analyse a SGL system, the Singular Isothermal Sphere (SIS) profile for the lens is the most frequently used density profile. In the SIS profile, the  total mass-density $\rho_{T}(r)$ is proportional to $ r^{-2}$. Recently, a power law model of the total mass-density profile has been studied. In this model, $\rho_{T} \propto r^{-\gamma}$, where $\gamma$ is the total mass density profile parameter and is also referred to  as the  power law index. 

Cao et al. considered these lens profiles and put constraints on cosmological parameters by keeping $\Omega_{m0}$ fixed  \cite{sc2015}. Study of the power law index $\gamma$ is of great interest as it helps in our understanding of  
  the structure of a galaxy. 
  Various observational datasets have been used  to study the evolution of $\gamma$ with redshift \cite{xl2016,jl2017}. 
  Recently, the largest sample of SGL ($161$ datapoints) was compiled by Chen et al. to study the effect of the lens mass model on the cosmological parameters assuming the $\Lambda$CDM model \cite{yc2019}. It is important to note that in all these studies, a flat universe was assumed. Xia et al.  used $118$ SGL systems to put constraints on the curvature parameter and the lens profile parameters using the distance sum rule \cite{jq2017}. 
Using the same method, the cosmological and lens profile parameters were constrained by several authors \cite{sr2015,zl2018,jz2018,bw2019}.\\

Following the same line of thought, we use the distance sum rule for \textit{two different purposes}. First, to constrain the cosmic curvature parameter and lens density profile parameters using SGL, SN Ia and Gamma Ray Bursts (GRBs) observations. Second, to study the behaviour of the distance duality parameter and  the cosmic curvature parameter using Time-Delay Distance (TDD) data.  In the first part, we use the extended power law lens profile  to constrain the cosmological and galaxy lens parameters using a model-independent approach, namely the distance sum rule. Recently Chen et al. also worked with the extended power law lens profile using SGL observations but they adopted a model dependent approach by assuming the  $\Lambda$CDM model \cite{yc2019}. Wang et al. have used $161$ datapoints and considered all three lens mass density profiles (SIS, power law and extended power law) with $1048$ supernovae (upto $z=2.3$) to calibrate the distances of the lens galaxy and source. The limitation of the SN Ia data used by them is that they could not include all the SGL datapoints in their analysis. In our analysis, we use SN Ia and GRBs data in order to include all $161$ datapoints of the SGL data upto $z=3.6$. In addition, previous work based on the distance sum rule, did not consider the evolution of the total mass density lens profile power index $(\gamma)$ as a function of redshift in the extended power lens profile. We believe that it is important to study the evolution of $\gamma$ with the redshift because any evolution in $\gamma$ with $z$ could indicate that in the growth of massive galaxies, dissipative processes have played an important role \cite{aj2012}. We attempt to modify the constraints on the cosmic curvature and lens profile parameters by taking into account all the above mentioned factors.\\

In the second part of the analysis, we use time delay distances. It is known that the sources in SGL systems such as quasars produce an  observable delay in the observed time between multiple images. Therefore, apart from the  distance ratio analysis, another important quantity in SGL systems is  ``Time Delay''.  The measurement of the time delay is highly dependent on the background cosmology and hence the cosmological parameters especially, the Hubble constant $(H_0)$, can be constrained using time-delay observations \cite{sc2017,vb2017,sb2019,ce2019,kc2019,aj2019,gc2019,mm2019}. We further extend our work with time-delay measurements by using the distance sum rule. For this, we use the H0LiCOW\footnote{\url{http://shsuyu.github.io/H0LiCOW/site/}} sample. In cosmology, the angular diameter distance $d_A(z)$ and the luminosity distance $d_L(z)$ are related by the  ``Cosmic Distance Duality Relation (CDDR)" as  $d_A(z)(1+z)^2=d_L(z)$ \cite{gf2007}. It has been claimed that the violation of CDDR may be a hint of new or exotic physics. Therefore, it is important to test the validity of CDDR with observational datasets. Violation of this relation has been checked in earlier work \cite{ba2004, jp2004, fd2006, rf2010, rn2011, ar2016, ar2017, sc2011, rn2012, rf2017, sr20166, hn20188, cz2018}. In their work, different data-sets have been used to check the redshift dependence of CDDR. For example using the JLA data,  it was found that there is a violation of CDDR at $1\sigma$\cite{xl2019}. Keeping this in mind, we also check the validity of this relation. Based on the distance sum rule, we put constraints on the cosmic curvature parameter and the distance duality parameter $\left(\eta(z) \equiv \frac{d_A(z)(1+z)^2}{d_L(z)}\right)$. We also check the evolution of the distance duality parameter with redshift.\\ 
 
The outline of the paper is as follows: In Section 2, we discuss the distance ratio and time-delay distance in Strong Gravitational Lensing systems. In Section 3, we describe the methodology and the details of datasets used in this paper. The analysis and results  are explained in Section 4. Discussion and conclusions are  presented in Section 5.\\

\section{Strong Gravitational Lensing}
According to the General Theory of Relativity, light rays passing near matter get bent due to the presence of gravity. The bending of light gives rise to multiple images of the source. This effect is known as Strong Gravitational Lensing \cite{rn1996, ss2010, ps2006}. The bending of light is  related to the mass distribution within the lens. 
\subsection{Distance Ratio}
The mass distribution of the lensing galaxy is commonly modeled as a Singular Isothermal Sphere (SIS) or a Singular Isothermal Ellipsoid (SIE) \cite{lv2006}.
Here we consider a more general and complex model of the lens, namely the Extended Power Law (EPL) model. This model allows us to consider the luminosity density profile to be  different from the total mass density profile. Therefore, it gives us the freedom to consider the effect  of dark matter on the mass distribution. We model
the total mass (luminous and dark-matter) density $(\rho(r))$ and luminous density $(\nu(r))$ distributions as power laws
\begin{equation}\label{eq:sl3}
\rho(r)=\rho_{0}\left(\dfrac{r}{r_{0}}\right)^{-\gamma}, \quad \nu(r)=\nu_{0}\left(\dfrac{r}{r_{0}}\right)^{-\delta}
\end{equation}
where $r$ represents the radial coordinate from the center of the lens galaxy and  $\gamma$ and $\delta$ are two free parameters. In addition  we also consider an  anisotropic three-dimensional dispersion of velocity, which allows for the possibility that the radial velocity dispersion $(\sigma_r)$ and the tangential velocity dispersion $(\sigma_\theta)$ may be different. Therefore, we define an anisotropy parameter, $\beta(r)=1-\sigma_\theta^2/\sigma_r^2$. Using  Eq. (\ref{eq:sl3}) and the anisotropy parameter $\beta(r)$ and applying the spherical Jeans equation, one can define the distance ratio \cite{ga2006}.
\begin{equation}\label{eq:sl4}
d_R\equiv\dfrac{d_{\mathrm{A}}^{^\mathrm{ls}}}{d_A^{^\mathrm{os}}}=\dfrac{c^{2} \theta_\mathrm{E}}{4 \pi \sigma_{0}^{2}}\left(\dfrac{\theta_{\mathrm{ap}}}{\theta_\mathrm{E}}\right)^{2-\gamma} \times f(\gamma, \delta, \beta)
\end{equation}
where
$$
{\small
\begin{aligned}
f(\gamma, \delta, \beta)=\frac{(2 \sqrt{\pi})(3-\delta)}{(\xi-2 \beta)(3-\xi)} &\times\left[\frac{\Gamma[(\xi-1) / 2]}{\Gamma(\xi / 2)}-  \beta \frac{\Gamma[(\xi+1) / 2]}{\Gamma[(\xi+2) / 2]}\right]\\ 
& \times\frac{\Gamma(\gamma / 2) \Gamma(\delta / 2)}{\Gamma\left[(\gamma-1) / 2\right] \Gamma[(\delta-1) / 2]}
\end{aligned}
}
$$
Here $\xi=\gamma+\delta-2$.  Using spectroscopic data, $\sigma_0$ the  observed velocity dispersion can be  related to $\sigma_{\mathrm{ap}}$ (velocity dispersions measured within apertures of arbitrary
sizes) via $\sigma_0=\sigma_{\mathrm{ap}}\left[\theta_{\mathrm{eff}} /\left(2 \theta_{\mathrm{ap}}\right)\right]^{-0.066}$ \cite{yc2019}. Here $\theta_\mathrm{E}$ and $\theta_\mathrm{ap}$ are the angular radii of the Einstein ring and circular aperture respectively while $\theta_{\text{eff}}$ is the effective angular radius of  the lens galaxy.
This extended lens profile  reduces to the SIS model for  $\gamma=\delta=2$ and $\beta=0$. In  earlier work, the anisotropy parameter $\beta(r)$ was always taken to be independent of $r$ \cite{as2006, lv2006}. For individual lensing systems one cannot determine $\beta$ independently. Therefore, based on the well-studied sample of nearby elliptical galaxies, we marginalise $\beta(r)$ using a Gaussian prior with $\beta=0.18\pm 0.13$ \cite{og2001}. A similar approach has  also  been adopted by various authors \cite{rg2008, js2009, sc2017, yc2019, bw2019}. 

We consider the same Gaussian prior on the $\beta$ parameter throughout the paper, i.e. $\beta=0.18\pm 0.13$. However, for the remaining parameters we consider a flat prior over the  range of interest. Furthermore, in order to include the redshift evolution of the total mass-density, we consider two parametrisation for $\gamma$, namely $\gamma_{I}(z)=\gamma_0+\gamma_1z$ and $\gamma_{II}(z)=\gamma_0+\gamma_1z/(1+z)$.\\

\subsection{Time-Delay Distance} 
Apart from distance ratio in SGL systems, time-delay is another important observation which can be used to put constraints on cosmological parameters in a model-independent way. The  light rays emitted at the same time from a source will reach the observer at different times as these paths have different path lengths and pass through different gravitational potentials. Therefore, there is a time-delay between the multiple images. If the source is a variable light source, this time-delay can be determined  by monitoring the images created by the lens which give us the flux information corresponding to the same source event. Time-delay is related to a quantity, called the ``time-delay distance",  which can be used to estimate cosmological parameters. The time-delay distance gives a relation between the three angular diameter distances, i.e.  observer-lens $ d_\mathrm{A}^{^\mathrm{o l}}$, lens-source $d_\mathrm{A}^{^\mathrm{l s}}$  and observer-source $d_\mathrm{A}^{^\mathrm{o s}}$ angular diameter distances. 
\vspace{2mm}\\
For a given source position ($\mathcal{B}$) and image position ($\theta$), the difference in time-delay $(\delta t)$ between the perturbed and unperturbed light rays is \cite{ps2006}
$$
\delta t(\theta, \mathcal{B})=\dfrac{\left(1+z_{l}\right)}{c} \dfrac{d_\mathrm{A}^{^\mathrm{o s}} d_\mathrm{A}^{^\mathrm{o l}}}{d_\mathrm{A}^{^\mathrm{l s}}}\left[\dfrac{(\theta-\mathcal{B})^{2}}{2}-\psi(\theta)\right]
$$
where $z_l$ is the lens redshift and $\psi$ is the effective gravitational potential of the lens. We can define a quantity, the time-delay distance, as 
$$d_{\Delta t}^{^{\mathrm {model }}} = \left(1+z_l\right)\frac{{d_\mathrm{A}^{^\mathrm{o s}} d_\mathrm{A}^{^\mathrm{o l}}}}{{d_\mathrm{A}^{^\mathrm{l s}}}}$$\\
In case of a two-image lens system, say image $i$ and $j$, the difference in time-delay between the two images $\Delta t_{i j}$ is given by 
\begin{equation}\label{eq:sl4a}
\begin{aligned}
\Delta t_{i j}\equiv \delta t_j-\delta t_i=\dfrac{d_{\Delta t}^{^{\mathrm {model }}}}{c}&\left[\dfrac{\left({\theta}_{j}-{\mathcal{B}}\right)^{2}}{2}-\psi\left({\theta}_{j}\right)\right.\\
&\left.-\dfrac{\left({\theta}_{i}-{\mathcal{B}}\right)^{2}}{2}+\psi\left({\theta}_{i}\right)\right]
\end{aligned}
\end{equation}

 The gravitational potential of the lens at image positions, $\psi\left({\theta}_{i}\right) \text { and } \psi\left({\theta}_{j}\right)$ and the source position $\mathcal{B}$ can be determined from the mass model of the system. From the measurement of $\Delta{t_{ij}}$ and by determining  $\psi(\theta)$ from an accurate model of a lens, it is possible to determine $d_{\Delta t}$ which can further be used to put bounds on the Hubble constant $(H_0)$ by assuming various  cosmological models. Thus, the time-delay distance method to put constraints on cosmological parameters 
  depends on three main ingredients
  \vspace{1mm}
\begin{itemize}[noitemsep,topsep=0pt]
\item 
A precise measurement of time-delay.
\item
The analysis of the combined lensing effect of all the mass distributions  along the  line of sight  up to the redshift of the lensed quasar.
\item
Well constrained models of nearby lens galaxies.
\end{itemize}
\vspace{1mm}
The mass of the lens galaxy that contributes to the potential of the lens galaxy plays a major role in the observed time delay. But in fact, the mass along the line of sight  between the observer and the source could  also cause focusing and defocusing of the rays and may impact the observed time delays \cite{us1994}. If this effect is not taken into account properly, one may get  biased $d_{\Delta t}$ inferences. We can include the contribution of this LOS interference effect by a parameter $ \kappa_{\mathrm{ext}}$ called the  external convergence in the lens plane. Assuming a small line of sight effect, we can approximate its effect on the time delay distance and $H_0$ as \cite{cr2003, cm2014}

\begin{equation}\label{eq:sl4b}
d_{\Delta t}=\dfrac{d_{\Delta t}^{^{\mathrm {model }}}}{1-\kappa_{\mathrm{ext}}}~~~~~\&~~~~~~H_{0}=\left(1-\kappa_{\mathrm{ext}}\right) H_{0}^{^{\mathrm{model}}}
\end{equation}
where the average value of $\kappa_{\mathrm{ext}}$ around the sky is zero. From Eq. (\ref{eq:sl4b}) we can see that $d_{\Delta t}$ can be biased because of  $\kappa_{\mathrm{ext}}$ and  hence this effect  needs to be taken into account. But constraining this external convergence from strong lens modeling alone is prevented by the mass sheet degeneracy \cite{ee1985}. An independent LOS structure modeling can predict the external convergence. 

In the time-delay analysis, we consider the recent dataset of H0LiCOW ($H_0$ Lenses in COSMOGRAIL’s Wellspring) collaboration which  measured the Hubble constant 
with  better accuracy using a sample of six time-delay lenses (B1608+656, RXJ1131-1231, HE 0435-1223, SDSS 1206+4332, WFI2033-4723, PG 1115+080). In their recent work, they constrained $H_0$ with $2.4\%$ precision assuming a flat $\Lambda$CDM model. The value of $H_0$ obtained by the H0LiCOW collaboration is in good agreement with the value obtained from  local measurements, i.e. type Ia supernovae \cite{kc2019}. In this sample of lenses, the primary ingredients for each of the lenses are obtained by observational follow-ups and innovative analytical methodologies. They used high-quality lensed quasar light curves, obtained primarily through optical monitoring by the COSMOGRAIL (COSmological MOnitoring of GRAvItational Lenses) project \cite{vb2017,ae2005}, radio-wavelength monitoring \cite{cd2002}, deep Hubble Space Telescope (HST) and/or ground-based adaptive optics (AO) imaging \cite{gc2019}, spectroscopy of the lens galaxy to measure its velocity dispersion \cite{dss2019}, and deep wide-field spectroscopy and imaging to characterize the LOS in these systems \cite{ce2019}.

\section{Methodology and Data Samples}
In this analysis we modify the Distance Sum Rule (DSR) method in two different forms in order to accommodate distance ratio and time-delay distance. For this, we use four  datasets, namely distance ratio and time delay distance in Strong Gravitational Lensing (SGL), Supernovae Ia (SN Ia) and Gamma Ray Bursts (GRBs).\\

\subsection{Distance Sum Rule Method}
Under the assumption of homogeneity and isotropy of the universe, one can define the dimensionless comoving distances $(D_\mathrm{co})$ as
 \begin{equation}\label{eq:sl7b}
\begin{aligned}
&D_\mathrm{co}^{^\mathrm{os}}\equiv D_\mathrm{co}(0,z_s)\equiv\dfrac{H_0}{c}d_\mathrm{co}^{^\mathrm{os}};\\
&D_\mathrm{co}^{^\mathrm{ol}}\equiv D_\mathrm{co}(0,z_l)\equiv\dfrac{H_0}{c}d_\mathrm{co}^{^\mathrm{ol}};\\
&D_\mathrm{co}^{^\mathrm{ls}}\equiv D_\mathrm{co}(z_l,z_s)\equiv\dfrac{H_0}{c}d_\mathrm{co}^{^\mathrm{ls}}
\end{aligned}
\end{equation}
where $d_\mathrm{co}^{^\mathrm{os}}$, $d_\mathrm{co}^{^\mathrm{ol}}$ and $d_\mathrm{co}^{^\mathrm{ls}}$ represent the comoving distances between observer-source, observer-lens and lens-source respectively.
 According to the Distance Sum Rule \cite{pj1993,sr2015}, the distance ratio, i.e. ratio of angular diameter distance at source and lens redshift to the angular diameter distance at source redshift is given by
\begin{equation}\label{eq:sl8}
\dfrac{D_{\mathrm{co}}^{^{\mathrm{ls}}}}{D_\mathrm{co}^{^{\mathrm{os}}}}=\sqrt{1+\Omega_{k0} \left(D_\mathrm{co}^{^\mathrm{ol}}\right)^{2}}-\dfrac{D_\mathrm{co}^{^\mathrm{ol}}}{D_\mathrm{co}^{^\mathrm{os}}} \sqrt{1+\Omega_{k0} \left(D_\mathrm{co}^{^\mathrm{os}}\right)^{2}}
\end{equation}
We can also write the distance sum rule in terms of the time-delay distance 
 \begin{equation}\label{eq:sl9}
\begin{aligned}
\dfrac{D_{\mathrm{co}}^{^{\mathrm{ol}}}D_{\mathrm{co}}^{^{\mathrm{os}}}}{D_{\mathrm{co}}^{^{\mathrm{ls}}}}=&\left[\dfrac{1}{D_{\mathrm{co}}^{^{\mathrm{ol}}}}\sqrt{1+\Omega_{k0} \left(D_{\mathrm{co}}^{^{\mathrm{ol}}}\right)^{2}}-\right.\\
&\left.\dfrac{1}{D_{\mathrm{co}}^{^{\mathrm{os}}}} \sqrt{1+\Omega_{k0} \left(D_{\mathrm{co}}^{^{\mathrm{os}}}\right)^{2}}\right]^{-1}
\end{aligned}
\end{equation}
The value of $\Omega_{k0}$ can thus be directly obtained from Eqs. (\ref{eq:sl8},\ref{eq:sl9}) without assuming any fiducial cosmological model if the distances $D_{\mathrm{co}}^{^{\mathrm{ol}}}$ and $D_{\mathrm{co}}^{^{\mathrm{os}}}$ are known from observations. Eqs. (\ref{eq:sl8}, \ref{eq:sl9}) represent theoretical constructions of the distance ratio and the time-delay distance respectively. In order to get the left hand sides of these two equations, we use SGL observations. 
\subsection{SGL Systems}
We use two different kinds of SGL observations: distance ratio and time-delay distance.
\vspace{2mm}\\
$\bullet$ \textbf{Distance ratio in SGL}
\vspace{2mm}\\
For the distance ratio, we use  a sample of  SGL systems \cite{yc2019}, which is a collection of  $5$ systems from the LSD survey \cite{lv20022, lv2003, tt2002, tt2004}, $26$ from SL2S \cite{aj2011, as2013, as2015}, $57$ from SLACS \cite{as20088, mw2009, mw2010}, $38$ from an extension of SLACS for the Masses survey\cite{ys2015, ys2017}, $21$ from BELLS \cite{jr2011} and $14$ from BELLS-GALLERY \cite{ys2016, ys20166}. After combining  the datapoints of these surveys, we get $161$ galaxy-scale strong lensing systems \cite{yc2019}. This sample includes  information of the lens redshift $(z_l)$, source redshift $(z_s)$,  Einstein radius $(\theta_\mathrm{E})$, velocity dispersion $(\sigma_{\mathrm{ap}})$ measured inside the circular aperture with angular radii $\theta_{ap}$, and the half-light angular radius of the lens galaxy $\theta_{\mathrm{eff}}$. The redshift range of the lenses is $0.0624 \leq z_{l} \leq 1.004$ and the source redshift range is $0.197 \leq z_{s} \leq 3.595$.
\vspace{2mm}\\
$\bullet$ \textbf{Time-Delay Distance in SGL} 
\vspace{2mm}\\
For the time-delay distance, we use $6$ quad-image SGL systems from the H0LiCOW collaboration. The individual lenses and their time-delay distances $(d_{\Delta t})$ are listed in Table \ref{tb:a1}. This dataset has been used for cosmological inferences in past \cite{kc2019,jj2020}. The main sources of the uncertainties in the estimation of the time-delay distances are the  time-delay measurement, LOS effect and the lens model. Out of these three uncertainties, the contribution from the estimation of the time delay and determination of the LOS are based on a Gaussian approximation and rest of the uncertainty comes from the lens model assumption as well as other unknown sources.

In order to minimize the error, the H0LiCOW collaboration considered only those data points that have small uncertainties in the above-mentioned three sources of error. For more details on the uncertainties that contribute to the errors,  see \cite{kc2019}.

\begin{table}[H]
\centering
\renewcommand{\arraystretch}{2}
     \begin{tabular}[b]{| c | c |c| c|}\hline
       Lens Name & $z_l$ & $z_s$ & $d_{\Delta t}(\mathrm{Mpc})$\\ \hline \hline
    B1608+656 & 0.6304 & 1.394        &$5156_{-236}^{+296}$  \\ \hline
    RXJ1131-1231 & 0.295 & 0.654   & $2096_{-83}^{+98}$ \\ \hline
    HE 0435-1223 & 0.4546 & 1.693   & $2707_{-168}^{+183}$ \\ \hline
    SDSS 1206+4332 & 0.745 & 1.789  & $5769_{-471}^{+589}$  \\ \hline
    WFI2033-4723 & 0.6575 & 1.662    &  $4784_{-248}^{+399}$\\ \hline
    PG 1115+080 & 0.311 & 1.722            & $1470_{-127}^{+137}$ \\ \hline
    \end{tabular}
\caption{ H0LiCOW sample. }
\label{tb:a1}
\end{table}

\subsection{Type Ia Supernovae}
We use the latest Pantheon dataset of type Ia supernovae to estimate the luminosity distance. The Pantheon dataset is the largest SN Ia sample till date having 1048 SN Ia in the redshift range  $0.01<z<2.26$ \cite{dm2018}. To determine the observed distance modulus, Scolnic et al.  \cite{dm2018} used the SALT2 \cite{jg2010} light curve fitter
$$
\mu_{\mathrm{SN}}=m_\mathrm{B}(z)+\alpha \cdot X_{1}-\beta \cdot \mathcal{C}-M_\mathrm{B}
$$
where $m_\mathrm{B}$ is the rest frame B-band peak magnitude, $M_\mathrm{B}$ represents absolute B-band magnitude of a fiducial SN Ia with $X_1 =0$ and $\mathcal{C}=0$, $X_1$ and $\mathcal{C}$ represent the time stretch of light curve and  supernova colour at maximum brightness respectively. The stretch-luminosity parameter $(\alpha)$ and the colour-luminosity parameter $(\beta)$ are calibrated to zero for the Pantheon sample, hence the observed distance modulus reduces to $\mu_{\mathrm{SN}}=m_\mathrm{B}-M_\mathrm{B}$. For a standard cosmological system, the distance modulus can be defined as
$$
\mu_{\mathrm{th}}=5 \log_{10} \left(d_\mathrm{L} / \mathrm{Mpc}\right)+25
$$
Thus, we estimate the luminosity distance $(d_\mathrm{L})$ and uncertainty in the luminosity distance $(\sigma_{d_\mathrm{L}})$ for each SN Ia as
 \begin{equation}\label{eq:sl10}
{ \small d_\mathrm{L}(z)=10^{\left(\mu_{\text{SN}}-25\right) / 5}~(\text{Mpc}),~~\sigma_{d_\mathrm{L}}=\dfrac{\ln(10)}{5}d_\mathrm{L}\sigma_{\mu_{\text{SN}}}~(\text{Mpc})}
\end{equation}
From Eq. (\ref{eq:sl10}) it is clear that the luminosity distance can be estimated by knowing the absolute magnitude of the supernovae $(M_\mathrm{B})$. Recent studies seem to indicate that there is no evolution of luminosity (or absolute magnitude) of Type Ia supernovae with redshift \cite{bm2020}. So, it is usually accepted that the Type Ia supernovae sample is normally distributed with a mean absolute magnitude of $M_\mathrm{B}=-19.22$. Therefore, we use $M_\mathrm{B}=-19.220 \pm 0.042$ to calculate the luminosity distance and its uncertainty for each supernova \cite{gb2020}. The supernova data that we use for the luminosity distance is up to $z=2.26$. However, the  SGL data is up to  $z=3.595$. Hence, in order to include all datapoints of the SGL data in our analysis, we look for another standard candle (GRBs) which could help us estimate the luminosity distance at higher redshifts.
\vspace{3mm}\\
\subsection{Gamma-Ray Bursts}
Gamma Ray Bursts (GRBs) are highly energetic events that occur in the universe and can be detected at a very high redshift due to their high luminosity. To date, the farthest GRB 090429B \cite{ac2011} observed is at $z=9.4$. GRBs are considered  an effective tool to study the universe \cite{li2015, hn2015, hn2016, jj2017}. Several efforts have been made to establish distance measures using some empirical relations of distance-dependent quantities and observables of rest frames \cite{la2008}. We consider the relation between the isotropic equivalent gamma-ray energy $E_{\gamma,\text{iso}}$ and the observed photon energy of the peak spectral flux $E_{\text{p,i}}$ \cite{la2002,la2006}
 \begin{equation}\label{eq:sl11}
\log \left(\dfrac{E_{\gamma,\text {iso }}}{1 \text { erg }}\right)=a \log \left[\dfrac{E_{\mathrm{p}, \mathrm{i}}}{300 \mathrm{keV}}\right] +b
\end{equation}
 $E_{\mathrm{p}, \mathrm{i}}=E_{\mathrm{p,obs}}(1+z)$ and $a$ and $b$ are, constants. $E_{\text {p,i }}$ and $E_{\mathrm{p,obs}}$ are the spectral peak energy in the cosmological rest frame of GRBs and in the observer's frame respectively. On the other hand, isotropic equivalent gamma-ray energy $E_{\gamma,\text{iso}}$ can be calculated as
\begin{equation}\label{eq:sl12}
E_{\gamma, \text { iso }}=\dfrac{4 \pi d_\mathrm{L}^{2}(z,p) S_{\mathrm{bolo}}}{(1+z)}
\end{equation}
where $S_{\mathrm{bolo}}$ is the bolometric gamma-ray fluency and $p$ represents the cosmological parameters.
From Eq. (\ref{eq:sl12}), we can calculate the luminosity distance for each GRB. To use GRBs as standard candles, this relation must be consistently calibrated \cite{mg2008,md2011, md2012, hg2012, sg2014, hn20155}. In this work, we use the latest GRB sample having 162 datapoints upto a redshift $9.4$  \cite{md2017}. However SGL data is  upto a  redshift of $3.6$. Therefore, we drop the GRBs which are above $z=3.6$. Hence we are left with  $147$ GRBs.\\

To summarize, DSR is modified to accommodate SGL observations such as distance ratio and time delay distance (See Eqs. (\ref{eq:sl8},\ref{eq:sl9})).  As  mentioned earlier, we need the luminosity distance corresponding to each SGL observation and for this, we use GRB data. In this dataset, the observables are $z$, $S_{\text{bolo}}$ and $E_{\mathrm{p}, \mathrm{i}}$. Since the luminosity distance $d_\mathrm{L}$ is related to 
$E_{\gamma, \text { iso }}$ (Eq. (\ref{eq:sl12})), we need to obtain $E_{\gamma, \text { iso }}$. This is obtained from Eq. (\ref{eq:sl11}) after calibration with SN Ia data \cite{md2017,fy2011,fy2015,js2016}. In order to match redshift of SGL observations and luminosity distance, we fit a second order polynomial\footnote{A higher order polynomial fit doesn't show a substantial deviation from the second order polynomial fit.} on the SN Ia and GRB data. For the fitting we use all the datapoints of the SN Ia data\footnote{We ignore off-diagonal terms in the covariance matrix  of the distance modulus and just focus on the statistical errors\cite{px2017}.} and only $147$ GRBs out of $162$ (upto a redshift of  $3.6$). The second order polynomial we use is 
$$
d_\mathrm{L}(z)=d_1z+d_2z^2
$$
where $d_1$ and $d_2$ are two free parameters and are fitted using a Python based module{\textbf{ {\texttt{lmfit}}}}\footnote{https://github.com/lmfit/lmfit-py/}. We find $d_1=4227.53\pm 16.15$ Mpc, $d_2=1996.29\pm 49.05$ Mpc and $\text{cov}(d_1,~d_2)=-0.725$. Figure \ref{fig:sl1} shows the fitting curve with the $1\sigma$ and $2\sigma$ regions along with a theoretical construction of luminosity distance based on the $\Lambda$CDM model. In order to show the correctness of Figure \ref{fig:sl1}, we independently constrain the free parameters of this second order polynomial using a Bayesian analysis based Python module {\textbf{ {\texttt{emcee}}}} and find the same best fit values of parameters.

\begin{widetext}
 \begin{center}
\begin{figure}[ht]
\centering
\centerline{\includegraphics[totalheight=8.3cm]{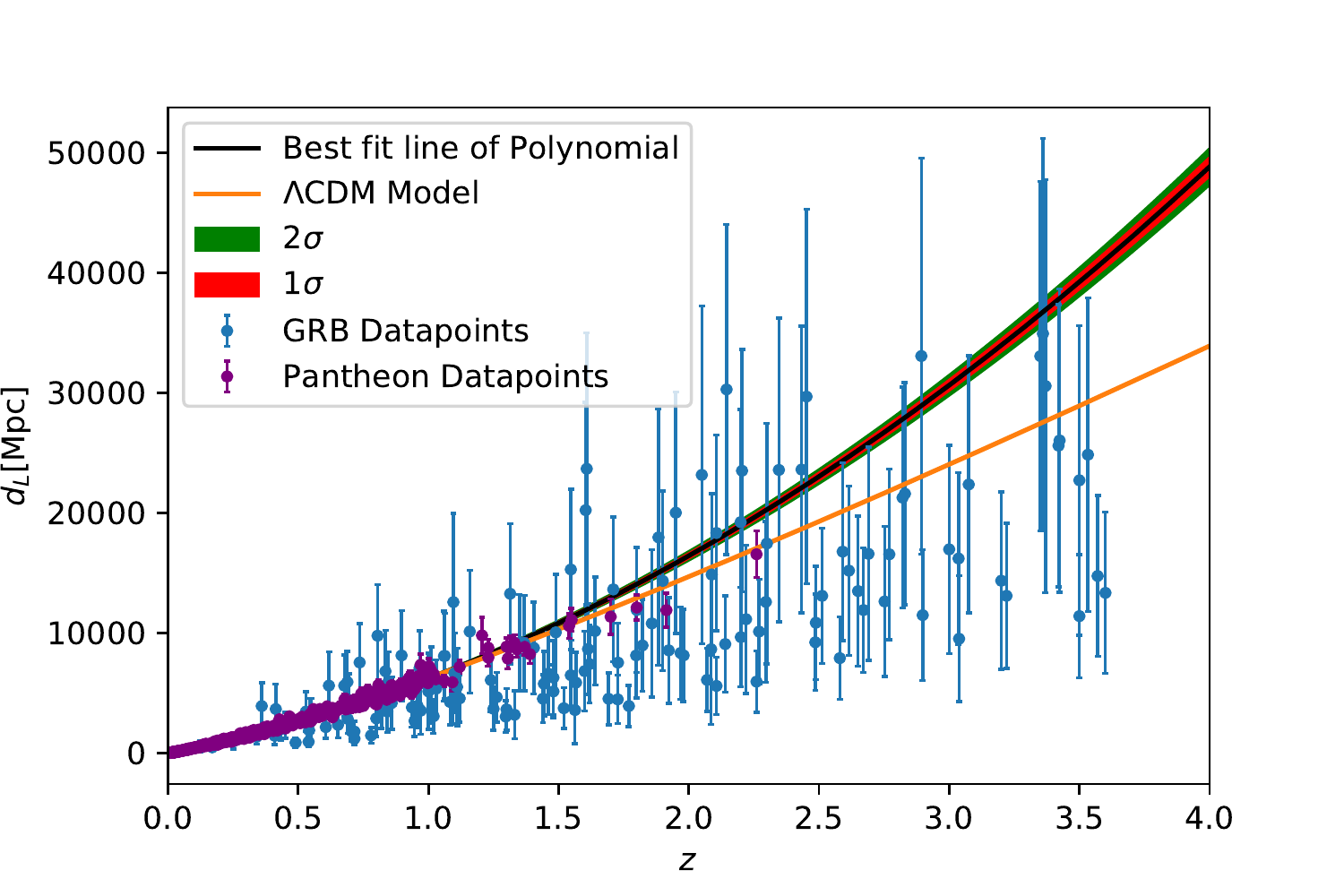}}
    \caption{Reconstruction of the Luminosity distance $d_L$ in Mpc from SN Ia and GRB datasets upto redshift $3.6$. The $68\%$ and $95\%$ confidence levels are represented by red and green shaded regions respectively. Violet and blue points with the error bars represent SN Ia and GRBs datapoints respectively. The solid orange line represents the luminosity distance for the $\Lambda$CDM model with $H_0=74.03$  $~\text{km sec}^{-1} \text{Mpc}^{-1}$. }
    \label{fig:sl1}
\end{figure}
 \end{center}
\end{widetext}

\section{Results}
The cosmological parameters and the lens profile model parameters are determined by maximising the likelihood $\mathcal{L} \sim \exp \left(-\chi^{2} / 2\right)$, where chi-square ($\chi ^2$) is
\begin{equation}\label{eq:sl14}
\chi^{2}\left(\mathbf{p_C}, \mathbf{p_L}\right)=\displaystyle\sum\limits_{i=1}^{n} \dfrac{\left(\mathcal{D}_{th}\left(z_{i} ; \mathbf{p_C}\right)-\mathcal{D}_{o b s}\left(z_{i} ; \mathbf{p_L}\right)\right)^{2}}{\sigma_{\mathcal{D}}\left(z_{i}\right)^{2}}
\end{equation}
Here $\mathbf{p_C}$ and $\mathbf{p_L}$ represent the cosmological parameters and the lens profile parameters respectively. $\mathcal{D}_{th}$ and $\mathcal{D}_{obs}$ are the theoretical and observed quantities of interest, i.e. the distance ratio and time-delay distance. Here $n$ stands for total number of datapoints used in the analysis. For distance ratio $n=161$ and for time-delay distance, $n=6$.\\

The two factors which contribute to the uncertainty in $\mathcal{D}$, i.e. $(\sigma_\mathcal{D})$ are the uncertainty in the observables of the SGL systems $(\sigma_{\text{SGL}})$ and uncertainty in the luminosity distance $(\sigma_{\text{SC}})$ (subscript ``SC'' stands for standard candles). We assume that the two uncertainties are uncorrelated and therefore they add in quadrature,  $\sigma_{\mathcal{D}}^{2}=\sigma_{S G L}^{2}+\sigma_{SC}^{2}$.
\vspace{3mm}\\
It is important to note that, for the validity of Eq. (\ref{eq:sl8}), the conditions 
$$
1+\Omega_{k0} \left(H_0d_\mathrm{L}^{^\mathrm{ol}}/c\right)^2 \geq 0 ~~~~~~\text{and}~~~~~~ 1+\Omega_{k0} \left(H_0d_\mathrm{L}^{^\mathrm{os}}/c\right)^2 \geq 0
$$
should hold. Therefore, based on the maximum luminosity distance,  we set a prior range of  cosmic curvature in our Markov Chain Monte Carlo (MCMC) program as $\Omega_{k0}>-0.2$. We also fix $H_0 = 74.03\pm 1.42 ~\text{km sec}^{-1} \text{Mpc}^{-1}$ observed from the Cepheid-supernova distance ladder \cite{ag2019} throughout our analysis.
\vspace{5mm}\\
\textbf{Distance Ratio: Constraint on Lens and Cosmological Parameters}
\vspace{1mm}\\
The Extended Power Law profile  is described by two power law indices- the power index of total mass density of a lens ($\gamma$ ) and the power index of the luminous density ($\delta$). In this analysis, we discuss two different parametrisations of $\gamma$ while $\delta$ is considered as a free parameter. The luminous density profile of the lens is different from the profile of total mass-density $(\gamma\neq\delta)$. 
\vspace{2mm}\\
Using Eqs. (\ref{eq:sl7b}, \ref{eq:sl8}), we can rewrite a theoretical distance ratio as
\begin{equation}\label{eq:sl12b}
\begin{aligned}
d_{\mathrm{R}}^{^{\mathrm{th}}}\equiv\dfrac{d_\mathrm{A}^{^{\mathrm{ls}}}}{d_\mathrm{A}^{^{\mathrm{os}}}}=&\sqrt{1+\Omega_{k 0}\left(\dfrac{ H_{0} d_{\mathrm{L}}^{^\mathrm{ol}}}{c\left(1+z_{l}\right)}\right)^{2}}-\dfrac{ d_{\mathrm{L}}^{^\mathrm{ol}}\left(1+z_{s}\right)}{d_{\mathrm{L}}^{^\mathrm{os}}\left(1+z_{l}\right)} \\
&\times\sqrt{1+\Omega_{k 0}\left(\dfrac{ H_{0} d_{\mathrm{L}}^{^\mathrm{os}}}{c\left(1+z_{s}\right)}\right)^{2}}
\end{aligned}
\end{equation}
The uncertainty in the theoretical distance ratio, i.e. $\sigma_{d_{\mathrm{R}}^{^{\mathrm{th}}}}$ can be calculated using the error propagation in Eq. (\ref{eq:sl12b}). On the other hand, the observed distance ratio $(d_{\mathrm{R}}^{^{\mathrm{obs}}})$ is defined in Eq. (\ref{eq:sl4}) and the corresponding uncertainty is calculated as 
\begin{equation}\label{eq:sl12c}
\sigma_{d_{\mathrm{R}}^{^{\mathrm{obs}}}}=d_\mathrm{R}^{^\mathrm{obs}}\sqrt{\left[\left(\dfrac{(\gamma-1)\sigma_{\theta_\mathrm{E}}}{\theta_\mathrm{E}}\right)^2+\left(\dfrac{2\sigma_{\sigma_0}}{\sigma_0}\right)^2\right]}
\end{equation}
We assume that these two uncertainties, i.e. $\sigma_{d_{\mathrm{R}}^{^{\mathrm{th}}}}$ and $\sigma_{d_{\mathrm{R}}^{^{\mathrm{obs}}}}$ are uncorrelated and therefore they add in quadrature; $\sigma_{d_{\mathrm{R}}}=\sqrt{\left(\sigma_{d_{\mathrm{R}}^{^{\mathrm{th}}}}\right)^2+\left(\sigma_{d_{\mathrm{R}}^{^{\mathrm{obs}}}}\right)^2}$. However, no error is assumed in ${\theta_{\mathrm{ap}}}$.
All the uncertainties have been tabulated in Table \ref{tb:udr}

\begin{table}[ht]
\caption{Uncertainties used in the distance ratio part}
\renewcommand{\arraystretch}{1.3} 
\centering
    \begin{tabular}[b]{| l | c | c | c |}\hline
       Parameter & Symbol & Uncertainty & Reference\\ \hline \hline 
    $\text{Einstein radius}^a$ & $\theta_E$ & $5\%$ & \cite{rf2016} \\ \hline
   $\text{Velocity dispersion}^b$ & $\sigma_0$ & as given in data & \cite{yc2019} \\ \hline
   Anisotropy parameter & $\beta$ & 0.13& \cite{og2001} \\ \hline
    \end{tabular}\\
    \footnotesize{$^a$ This uncertainty value is taken to be the same for all lens systems.\hspace*{3cm} \\$^b$ The uncertainty is 
    	different for each lens system and is taken from the data.\hspace*{1.20cm}}
\label{tb:udr}
\end{table}
\vspace{1mm}
\begin{flushleft}
\textbf{P1:~}${{\gamma_{I}(z)=\gamma_0+\gamma_1z_l}}$
\end{flushleft}
\vspace{2mm}
In the first parametrisation, we consider $\gamma$ as a function of the redshift. The best fit values of $\Omega_{k0}$ and lens profile parameters are given in Table \ref{tb:sl4}.
\begin{table}[ht]
\centering
\renewcommand{\arraystretch}{2}
\begin{tabular}[b]{| c | c |} \hline
       Parameter & Best value [68\% C.L.] \\ \hline \hline
    $\Omega_{k0}$ & $-0.004^{+0.184}_{-0.118}$ \\ \hline
   $\gamma_0$ & $2.154^{+0.043}_{-0.034}$ \\ \hline
   $\gamma_1$ & $-0.037^{+0.075}_{-0.094}$ \\ \hline
   $\delta$ & $2.108^{+0.221}_{-0.325}$ \\ \hline
    \end{tabular}
    \caption{The best fit values of $\Omega_{k0},~\gamma_0,~\gamma_1$ and $\delta$ with $68\%$ confidence level for P1 parametrisation of EPL Model.}
    \label{tb:sl4}
\end{table}

\begin{figure}[h!]
\centering
 \epsfig{figure=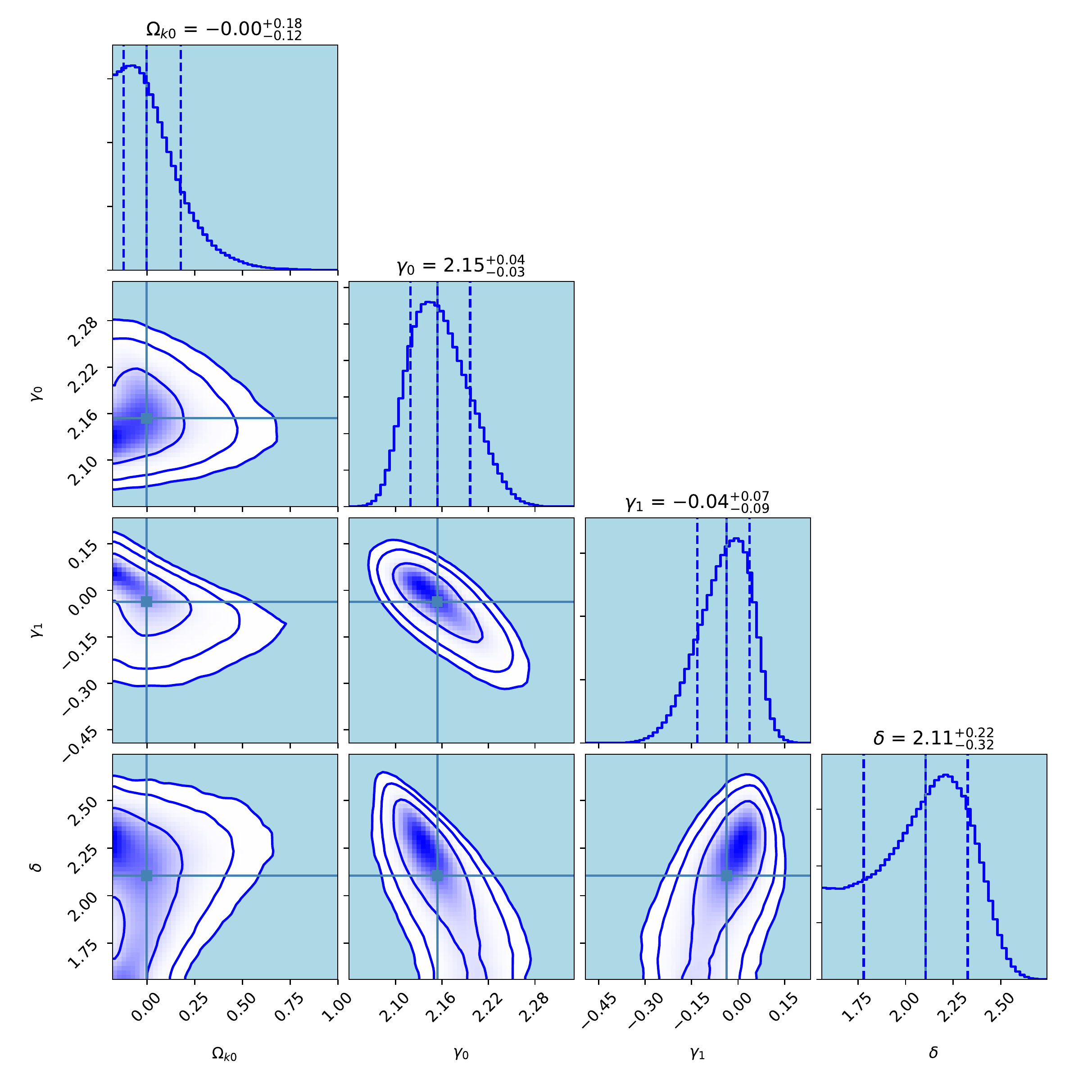,height=8cm,width=8cm,angle=0}
\captionof{figure}{$1D$ and $2D$ posterior distributions of $\Omega_{k0},~ \gamma_0,~ \gamma_1$ and $\delta$ for P1 parametrisation of EPL Model.}
\label{fig:sl5}
\end{figure}

The best fit value of $\Omega_{k0}$ is $-0.004^{+0.184}_{-0.118}$, which suggests that a spatially flat universe can be accommodated at $68\%$ confidence level. 
The estimated values of total mass density and luminous density profile of the lens are different. Further the results show a mild evolution of the total mass-density power index with redshift. The 1D and 2D posterior distributions of $\Omega_{k0}$, $\gamma_0,~\gamma_1$ and $\delta$ are shown in Figure \ref{fig:sl5}.

\begin{flushleft}
The lens density profile parameters, i.e. $\gamma_0$ and $\gamma_1$ show negative correlation as shown in figure \ref{fig:sl5}.
\end{flushleft}
\vspace{5mm}
\textbf{P2:~}${{\gamma_{II}(z)=\gamma_0+\gamma_1\dfrac{z_l}{1+z_l}}}$
\vspace{2mm}\\
In this parametrisation, we consider $\gamma$ as a function of redshift which converges to $\gamma_0$ at high redshift. The best fit values of $\Omega_{k0}$ and lens profile parameters are given in Table \ref{tb:sl5}.

\begin{table}[h]
\centering
\renewcommand{\arraystretch}{2}
\begin{tabular}[b]{| c | c |} \hline
       Parameter & Best value [68\% C.L.] \\ \hline \hline
     $\Omega_{k0}$ & $-0.032^{+0.168}_{-0.104}$ \\ \hline
   $\gamma_0$ & $2.163^{+0.066}_{-0.052}$ \\ \hline
   $\gamma_1$ & $-0.083^{+0.184}_{-0.243}$ \\ \hline
   $\delta$ & $2.064^{+0.265}_{-0.353}$ \\ \hline
    \end{tabular}
    \caption{The best fit values of $\Omega_{k0},~\gamma_0,~\gamma_1$ and $\delta$ with $68\%$ confidence level for P2 parametrisation of EPL Model.}
    \label{tb:sl5}
    \end{table}
    
\begin{figure}[h!]
\centering
 \epsfig{figure=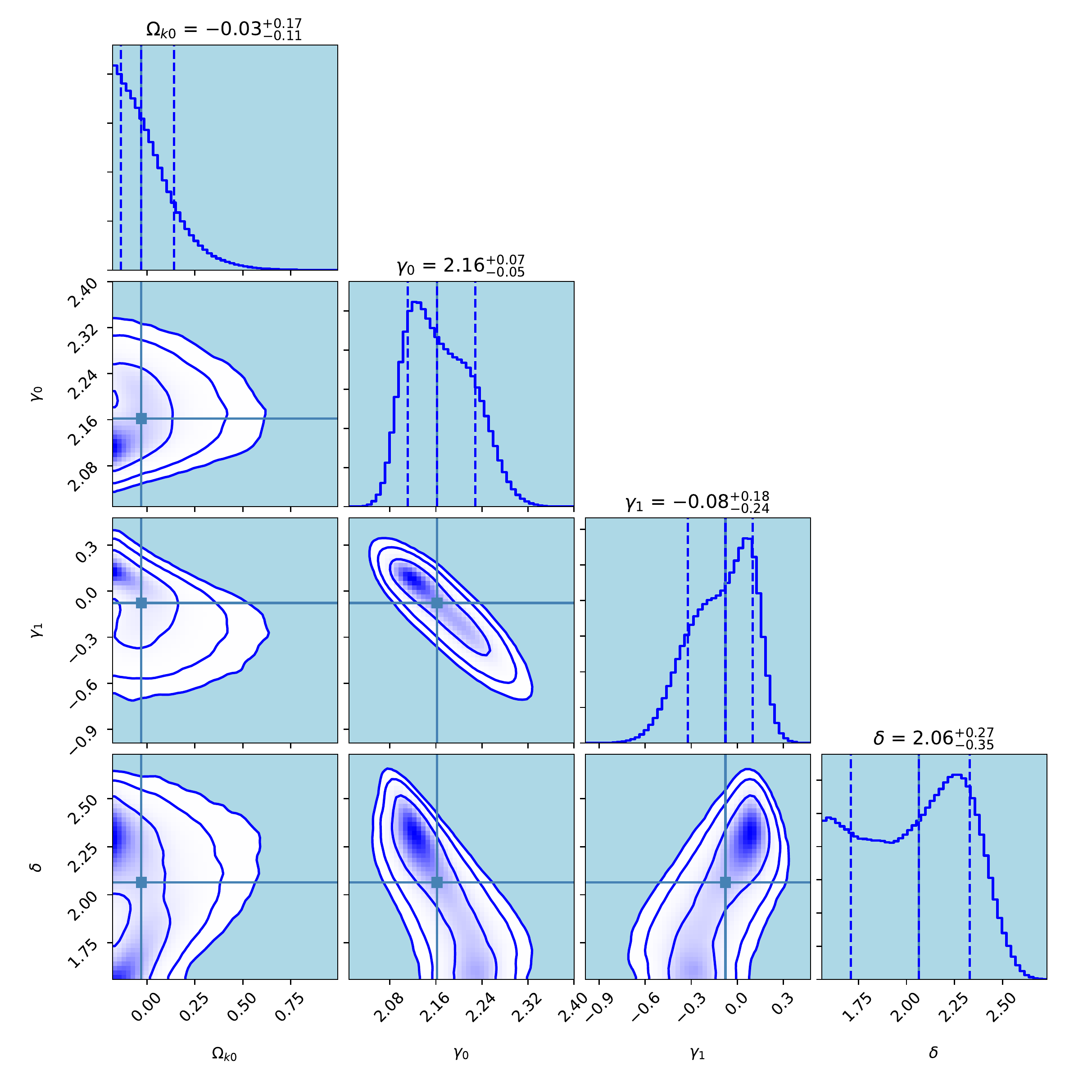,height=8cm,width=8cm,angle=0}
\captionof{figure}{$1D$ and $2D$ posterior distributions of $\Omega_{k0},~ \gamma_0,~ \gamma_1$ and $\delta$ for P2 parametrisation of EPL Model.}
\label{fig:sl6}
\end{figure}
The best fit value of $\Omega_{k0}$ is $-0.032^{+0.168}_{-0.104}$ and at $68\%$ confidence level, it again suggests a spatially flat universe. The different values of total mass density and luminous density seem to indicate a mild evolution of the total mass density power index with redshift. In Figure \ref{fig:sl6}, the 1D and 2D posterior distributions of $\Omega_{k0}$, $\gamma_0,~\gamma_1$ and $\delta$ are shown. This figure shows negative correlation between lens density profile parameters $\gamma_0$ and $\gamma_1$.
\vspace{2mm}\\
For the sake of completeness, we also did the same analysis with the Singular Isothermal Sphere \textbf{SIS} model and Power-Law Spherical \textbf{PLS} model. 
\vspace{5mm}\\
For SIS model, the distance ratio is
$$
\dfrac{d_{A}^{^{\mathrm{ls}}}}{d_{A}^{^{\mathrm{os}}}}=\dfrac{c^{2} \theta_{E}}{4 \pi f_{e}^{2} \sigma_{0}^{2}}
$$
where $f_e$ is a free parameter which quantifies the velocity dispersion due to the total mass of the lens (including dark matter) to the observed velocity dispersion of stars $\sigma_{0}$.\\
The distance ratio for the PLS model is 
$$
\dfrac{d_{A}^{^{\mathrm{ls}}}}{d_{A}^{^{\mathrm{os}}}}=\dfrac{c^{2} \theta_{E}}{4 \pi \sigma_{a p}^{2}}\left(\dfrac{\theta_{a p}}{\theta_{E}}\right)^{2-\gamma} f^{-1}(\gamma)
$$
where, \\
$f(\gamma)=-\dfrac{1}{\sqrt{\pi}} \dfrac{(5-2 \gamma)(1-\gamma)}{3-\gamma} \dfrac{\Gamma(\gamma-1)}{\Gamma(\gamma-3 / 2)}\left[\dfrac{\Gamma(\gamma / 2-1 / 2)}{\Gamma(\gamma / 2)}\right]^{2}$. We choose the redshift dependent form of the power law index $\gamma=\gamma_0+\gamma_1z_l$.
\vspace{2mm}\\
The best fit value of the parameters for SIS model and PLS model are given in Table \ref{tb:sl1111} and Table \ref{tb:sl2222} respectively.

\begin{table}[ht]
\centering
\renewcommand{\arraystretch}{2}
\begin{tabular}[b]{| c | c |} \hline
       Parameter & Best value [68\% C.L.] \\ \hline \hline
     $\Omega_{k0}$ & $0.680^{+0.144}_{-0.136}$ \\ \hline
   $f_e$ & $1.034^{+0.006}_{-0.006}$ \\ \hline
    \end{tabular}
\caption{The best fit values of $\Omega_{k0}$ and $f_e$ with $68\%$ confidence level for the SIS Model. }
\label{tb:sl1111}
\end{table}

\begin{table}[ht]
\centering
\renewcommand{\arraystretch}{2}
\begin{tabular}[b]{| c | c |} \hline
       Parameter & Best value [68\% C.L.] \\ \hline \hline
     $\Omega_{k0}$ & $-0.052^{+0.054}_{-0.050}$ \\ \hline
   $\gamma_0$ & $2.107^{+0.018}_{-0.020}$ \\ \hline
   $\gamma_1$ & $-0.371^{+0.088}_{-0.062}$ \\ \hline
    \end{tabular}
\caption{The best fit values of $\Omega_{k0}$, $\gamma_0$ and $\gamma_1$ with $68\%$ confidence level for the PLS Model. }
\label{tb:sl2222}
\end{table}
\newpage
\begin{flushleft}
\textbf{Time-Delay Distance: Constraint on Distance Duality \& Cosmological Parameters }
\end{flushleft}
\vspace{1mm}
The cosmic distance duality relation is one of the important concepts in cosmology. This is a relation between the luminosity distance and angular diameter distance. It is parametrised by the distance duality parameter $\eta(z)$
\begin{equation}\label{eq:sl15}
\eta(z)=\dfrac{d_\mathrm{A}(z)(1+z)^2}{d_\mathrm{L}(z)}
\end{equation}
\vspace{2mm}\\
The observed time-delay distance is defined as
\begin{equation}\label{eq:a1}
d_{\Delta t}^{^{\mathrm{obs}}}=\dfrac{(1+z_l)}{1-\kappa_{\mathrm{ext}}}\dfrac{d_{\mathrm{A}}^{^{\mathrm{ol}}} d_{\mathrm{A}}^{^{\mathrm{os}}}}{d_{\mathrm{A}}^{^{\mathrm{ls}}}}
\end{equation}
while the theoretical construction of time-delay distance $d_{\Delta t}^{\mathrm{th}}$ is given as
\begin{equation}\label{eq:a2}
\begin{aligned}
 d^{^{\mathrm{th}}}_{\Delta t}=\left[\dfrac{\left(1+z_{l}\right)}{\eta_{l} d_\mathrm{L}^{^\mathrm{o l}}} \right.&\left.\sqrt{1+\Omega_{k0}\left(\dfrac{\eta_{l} H_{0} d_\mathrm{L}^{^\mathrm{o l}}}{c\left(1+z_{l}\right)}\right)^{2}}-\dfrac{\left(1+z_{s}\right)}{\eta_{s} d_\mathrm{L}^{^\mathrm{o s}}} \right.\\
&\left.\times\sqrt{1+\Omega_{k0}\left(\dfrac{\eta_{s} H_{0} d_\mathrm{L}^{^ \mathrm{os}}}{c\left(1+z_{s}\right)}\right)^{2}}\right]^{-1}
 \end{aligned}
\end{equation}
where $\eta_l$ and $\eta_s$ are the distance duality parameters at lens and source redshift respectively.\\
Once we have the observed and theoretical time-delay distances, we can constrain the cosmic curvature $(\Omega_{k0})$ and distance duality $(\eta)$ parameter by maximising the likelihood $\mathcal{L} \sim \exp \left(-\chi^{2} / 2\right), \text { where chi-square }\left(\chi^{2}\right) \text { is }$
$$
\chi^{2}\left(\Omega_{k0},\eta\right)=\sum_{i=1}^{6} \dfrac{\left(d^{\mathrm{th}}_{\Delta t}\left(z_{i} ; \Omega_{k0},\eta\right)-d^{\mathrm{obs}}_{\Delta t}\left(z_{i} \right)\right)^{2}}{\sigma_{{d_{\Delta t}}}\left(z_{i}\right)^{2}}
$$
 $\sigma_{{d_{\Delta t}}}\left(z_{i}\right)$ is obtained on adding the uncertainties in the observables of the six lens system and luminosity distance in quadrature.
\vspace{2mm}\\
The results for the two parametrisations using the H0LiCOW sample are discussed below.
\vspace{2mm}
\begin{flushleft}
$\mathbf{P} \mathbf{1}: \eta_{I}=1+\eta_{1}z$
\end{flushleft}
In the first parametrisation of the distance duality parameter, we choose $\eta_{I}(z)=1+\eta_1z$. Table \ref{tb:a3} displays the best fit values of the cosmic curvature parameter $(\Omega_{k0})$ and distance duality parametrisation coefficient $(\eta_1)$ for the first parametrisation.

\begin{table}[ht]
\centering
\renewcommand{\arraystretch}{2}
\begin{tabular}[b]{| c | c |} \hline
       Parameter & Best value [68\% C.L.] \\ \hline \hline
    $\Omega_{k0}$ & $0.313^{+0.168}_{-0.196}$ \\ \hline
   $\eta_1$ & $0.249^{+0.173}_{-0.130}$ \\ \hline
    \end{tabular}
    \caption{The best fit values of $\Omega_{k0}$ and $\eta_1$ with $68\%$ confidence level for P1 parametrisation.}
    \label{tb:a3}
\end{table}

\begin{figure}[h!]
\centering
 \epsfig{figure=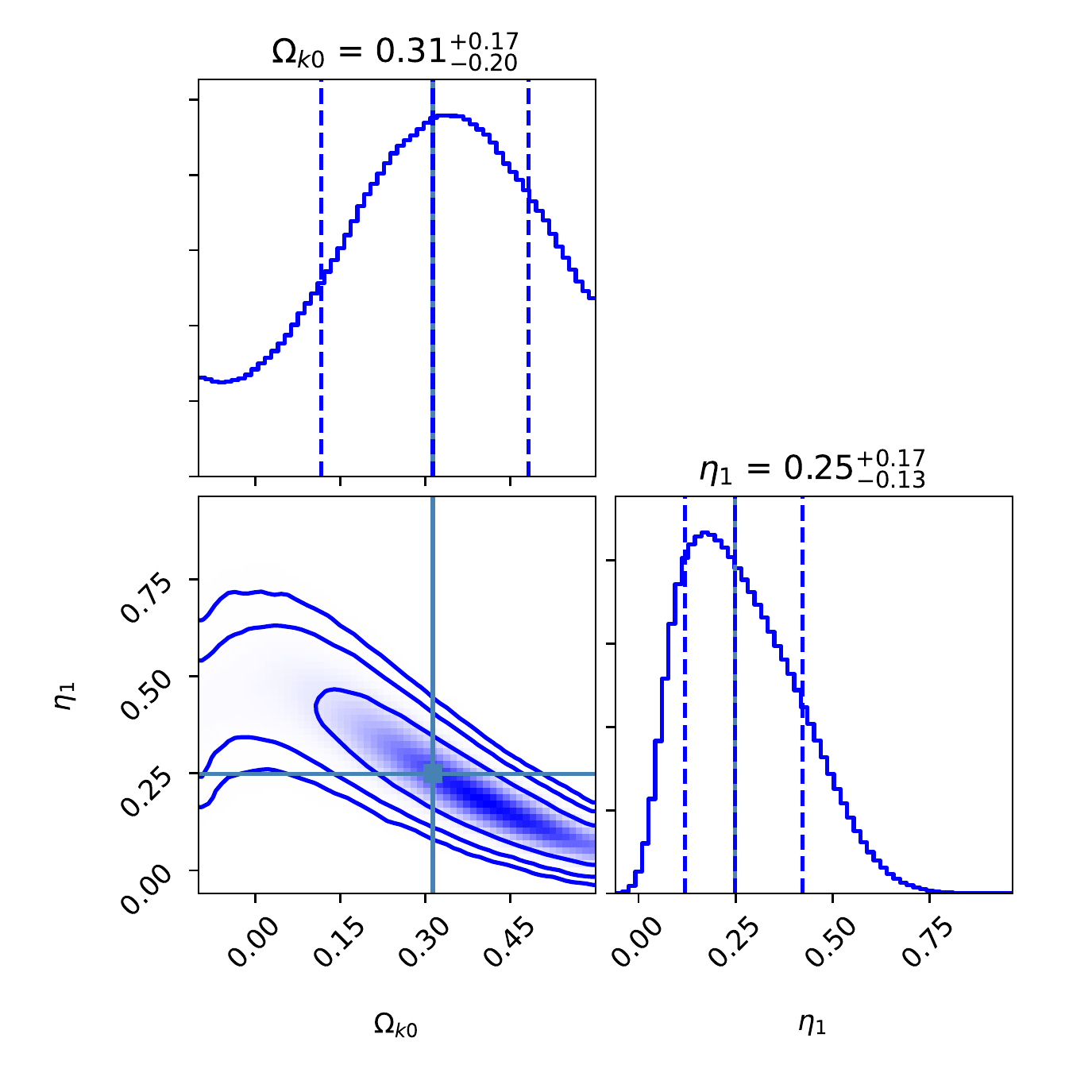,height=8cm,width=8cm,angle=0}
\captionof{figure}{$1D$ and $2D$ posterior distributions of $\Omega_{k0}$ and $\eta_1$ for P1 parametrisation.}
\label{fig:a2}
\end{figure}

The value of $\Omega_{k0}=0.313^{+0.168}_{-0.196}$ is in concordance with an open universe at $68\%$ confidence level and at $95\%$ confidence level a flat universe can also be accommodated. Further, $\eta_{1} \sim 0$  at $95\%$ confidence level shows no violation of the distance duality relation. The 1D and 2D posterior distributions of $\Omega_{k 0}$ and $\eta_1$ are shown in Figure \ref{fig:a2}.

\begin{flushleft}
$\mathbf{P} \mathbf{2}: \eta_{II}=1+\eta_{1}\dfrac{z}{1+z}$
\end{flushleft}
The value of best fit parameters for the second parametrisation are displayed in Table \ref{tb:a4}

\begin{table}[ht]
\centering
\renewcommand{\arraystretch}{2}
\begin{tabular}[b]{| c | c |} \hline
       Parameter & Best value [68\% C.L.] \\ \hline \hline
    $\Omega_{k0}$ & $0.113^{+0.197}_{-0.144}$ \\ \hline
   $\eta_1$ & $0.344^{+0.195}_{-0.185}$ \\ \hline
    \end{tabular}
    \caption{The best fit values of $\Omega_{k0}$ and $\eta_1$ with $68\%$ confidence level for P2 parametrisation.}
    \label{tb:a4}
\end{table}

\begin{figure}[h!]
\centering
 \epsfig{figure=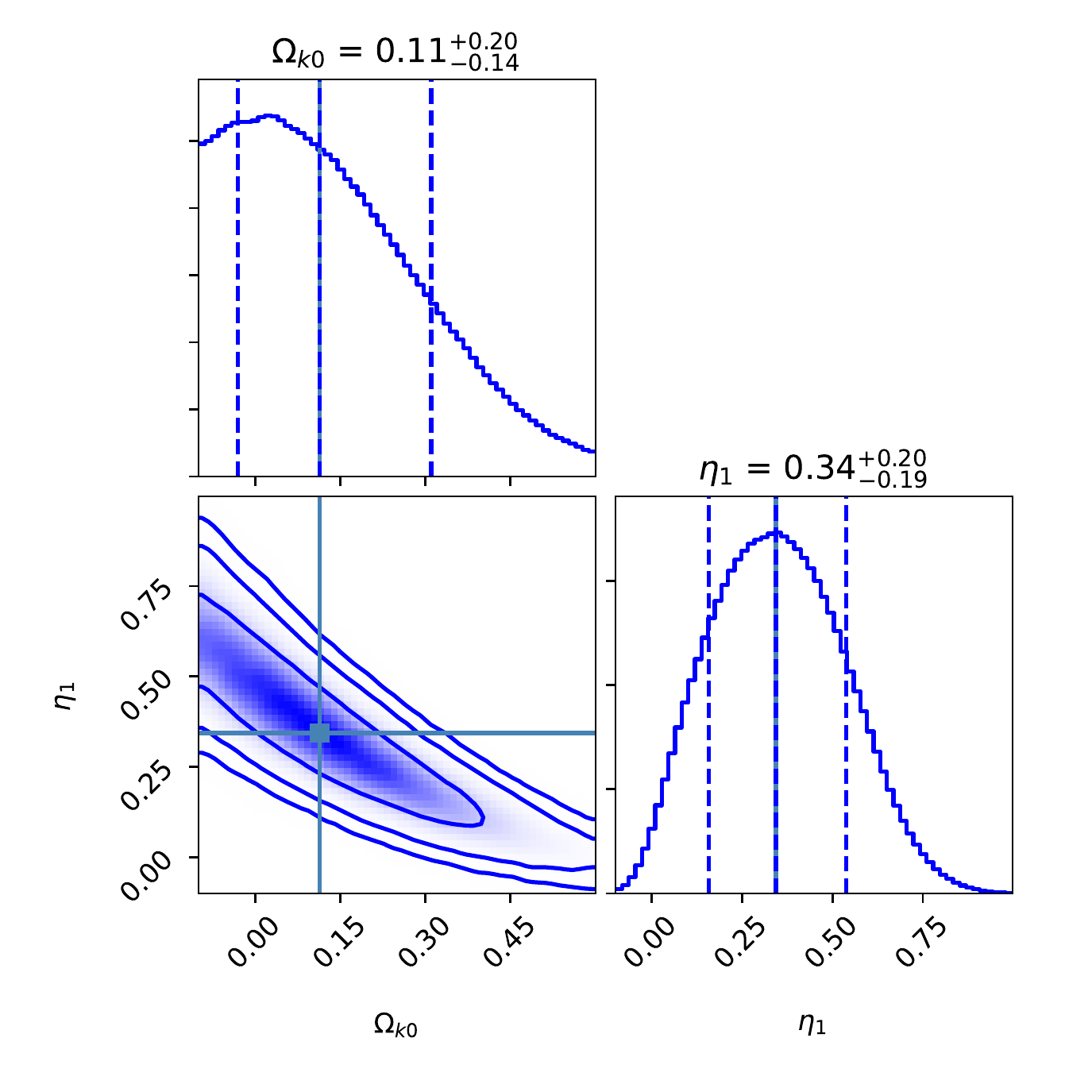,height=8cm,width=8cm,angle=0}
\captionof{figure}{$1D$ and $2D$ posterior distributions of $\Omega_{k0}$ and $\eta_1$ for P2 parametrisation.}
\label{fig:a3}
\end{figure}
The best fit value of $\Omega_{k0}$ indicates a mild preference for an open universe, and it is also in agreement with a flat universe at $68\%$ confidence level. Further, we find no violation of the distance duality relation with $\eta_{1} \sim 0$  at $95\%$ confidence level. We show the 1D and 2D posterior distributions of $\Omega_{k 0}$ and $\eta_1$ in Figure \ref{fig:a3}.

In the P1 and P2 parametrisations, a non-zero value of $\eta_1$ indicates a redshift evolution of the distance duality parameter. The 2D posterior plots for two cases of distance duality parameter show a correlation between cosmic curvature and distance duality parameters.

\section{Discussion and Conclusions}\label{sec:dc}
The Distance Sum Rule (DSR) along with strong gravitational lens systems is a powerful astrophysical tool to probe the curvature of the universe and galaxy parameters without assuming any fiducial cosmological model. We use DSR in two different ways. In the first part, we apply the DSR method with distance ratio introduced by R$\ddot{\textnormal{a}}\textnormal{s}\ddot{\textnormal{a}}$nen et al.  to measure  the cosmic curvature parameter along with the galaxy parameters in a model independent way \cite{sr2015}. In the second part, we again apply the DSR method to study the Cosmic Distance Duality Relation (CDDR) using time-delay distances data. We separately discuss the results of both the approaches.

\subsection{Method I: Distance Ratio}
In earlier studies, the distance ratio obtained from SGL was used in a model-dependent way to constrain different cosmological and lens parameters \cite{jl2017,sc2016, yc2019}. However, using the distance sum rule,  the distance ratio is used to constrain the cosmic curvature parameter along with lens galaxy parameters in a model-independent manner.  \cite{jq2017, zl2018, jz2018, bw2019}. Using the same methodology, we consider the latest SGL sample containing $161$ datapoints for the distance ratio part.\textit{ To include the full dataset of SGL systems in our analysis, we reconstruct the distance-redshift relation by including the SN Ia and GRBs data and  fit a second-order polynomial to  obtain the luminosity distance at the lens and source redshift of the SGL systems.} 
Further, to explore the nature of the lens (galaxy) profile,  we consider the  Extended Power Law lens model which allows us to use different profiles of the total mass density and luminous mass density of the lens. \textit{For this lens profile, we extend our work by considering the evolution of total mass density power index with redshift, something we believe has not been done in the EPL model.} \\ 

 Our main conclusions are listed below-
\begin{itemize}
\item
For completeness, we use DSR along with distance ratio data with the Singular Isothermal Sphere (SIS) and the  Power Law Spherical (PLS) lens profiles. For  the SIS profile, we obtain $\Omega_{k 0}=0.680_{-0.136}^{+0.144}$. Using the same lens profile Wang et al. obtained $\Omega_{k 0}=0.39_{-0.30}^{+0.22}$ \cite{bw2019}. Thus our results are in concordance with the results of Wang et al. but incompatible with the Planck result \cite{planck2018}. Though the SIS profile is one of the most frequently used lens profiles, the inconsistency of our results with the Planck result could be an indication that one should also explore other lens profiles that can provide better results with the distance ratio data.
For the PLS lens profile we find the constraint on the cosmic curvature parameter is consistent with a flat universe at $95\%$ confidence level. 
\item
 Extended Power Law (EPL) lens model along with DSR method involves three parameters: cosmic curvature parameter and two lens parameters, $\gamma$ (power index of total mass-density lens profile) and $\delta$ (power index of luminous density lens profile). We consider two different parametrisations of $\gamma$.  We assume the evolution of $\gamma$ with redshift in two different forms: $\gamma_{I}(z_l)=\gamma_0+\gamma_1z_l$, and $\gamma_{II}(z_l)=\gamma_0+\gamma_1z_l/(1+z_l)$. Both parametrisations of $\gamma$ indicate that there is a marginal evolution of $\gamma$ with redshift. For early galaxies, $\gamma(z)$ and $\delta$ are not identical. This might indicate that the distribution of dark matter and baryonic matter is not the same. In both the parametrisations of $\gamma$, the best fit values of $\Omega_{k0}$ indicate a closed universe but a spatially flat universe is also accommodated at $68\%$ confidence level. For both $\gamma_{I}$ and $\gamma_{II}$, the posterior distribution contours of cosmic curvature and lens parameters are very similar, suggesting that limits on the curvature parameter are not significantly affected by the choice of parametrisation of $\gamma$.

\item
The $1D$ and $2D$ posterior contour plots in curvature and lens profile parameter space indicate a strong correlation between them. We also find that the parameters of the lens profile are correlated among themselves.

\end{itemize}

\subsection{Method II: Time-Delay Distance}
 In the second part of the analysis, we test the validity of the Cosmic Distance Duality Relation (CDDR) based on the DSR method. Any strong evidence of the violation of CDDR could hint at the emergence of new physics. In the past, CDDR has already been verified using a variety of methods\cite{ba2004, jp2004, fd2006, rf2010, rn2011, ar2016, ar2017, sc2011, rn2012, rf2017, sr20166, hn20188, cz2018}. {\textit{In earlier work, DSR had been used with distance ratio. However, we believe this is the first time the DSR method has been modified to accommodate time-delay distance data in order to check the validity of CDDR.}} In this work, we use this method to put bounds on the cosmic curvature along with CDDR considering 6 datapoints of H0LiCOW samples of time-delay distance. We also consider redshift evolution of the distance duality parameter.  \\
A brief summary of the results is as follows:\\

\begin{itemize}[noitemsep,topsep=3pt]
\item
For the first parametrisation, the obtained values of $\Omega_{k0}$ and $\eta_1$ are $0.313_{-0.196}^{+0.168}$ and $0.249_{-0.130}^{+0.173}$ respectively. The  best fit value of cosmic curvature indicates an open universe at $68\%$ confidence level but at $95\%$ confidence level a flat universe is accommodated. The obtained value of distance duality parameter suggests that there is no violation in distance duality relation at $95\%$ confidence level.

\item
In the second parametrisation, the best fit  value of $\Omega_{k0}$ again suggests an open universe and a flat universe is accommodated at $68\%$ confidence level. Further, we find no violation in the distance duality parameter at $95\%$ confidence level.

\end{itemize}

\subsection{Diagonal Test: $d_{\Delta t}$ (flat $\Lambda$CDM) Vs $d_{\Delta t}$ (obs.) }
In order to compare the time-delay distances, $d_{\Delta t}$, obtained from observations and those predicted by a flat $\Lambda$CDM model, we apply a diagonal test on the H0LiCOW sample. This test highlights whether the two estimates are consistent or not. The analysis is as follows\\

We have tabulated the time-delay distances observed from the H0LiCOW sample and based on the flat $\Lambda$CDM model in Table \ref{tb:a6}. For the time-delay distances based on the flat $\Lambda$CDM model, we adopt the latest published value of $\Omega_{m 0}=0.3111 \pm 0.0073$ and $H_{0}=67.66 \pm 0.42 \mathrm{~km} \mathrm{~sec}^{-1} \mathrm{~Mpc}^{-1}$\cite{planck2018}.
\begin{table}[ht]
\centering
\renewcommand{\arraystretch}{2}
     \begin{tabular}{|c|c|c|c|c|}\hline  Sr. No.& $z_l$ & $z_s$ & $d_{\Delta t}^{\mathrm{obs}}$ [Mpc]  & $d_{\Delta t}^{\Lambda CDM}$ [Mpc] \\ \hline 
1 & 0.6304 & 1.394 &$5156_{-236}^{+296}$        &5337.079 \\ \hline 
2 & 0.2950 & 0.654& $2096_{-83}^{+98}$            & 2427.221 \\ \hline 
3 & 0.4546 & 1.693 &$2707_{-168}^{+183}$            & 2865.103 \\ \hline 
4 & 0.7450 & 1.789 &$5769_{-471}^{+589}$           & 5976.249 \\ \hline 
5 & 0.6575 & 1.662&  $4784_{-248}^{+399}$          & 5066.860 \\ \hline 
6 & 0.3110 & 1.722      & $1470_{-127}^{+137}$       & 1734.135 \\ \hline
\end{tabular}
\caption{Time-delay distance comparison between the observed sample and a theoretical prediction based on flat $\Lambda$CDM model. }
\label{tb:a6}
\end{table}

\begin{figure}[h!]
    \centering
     \epsfig{figure=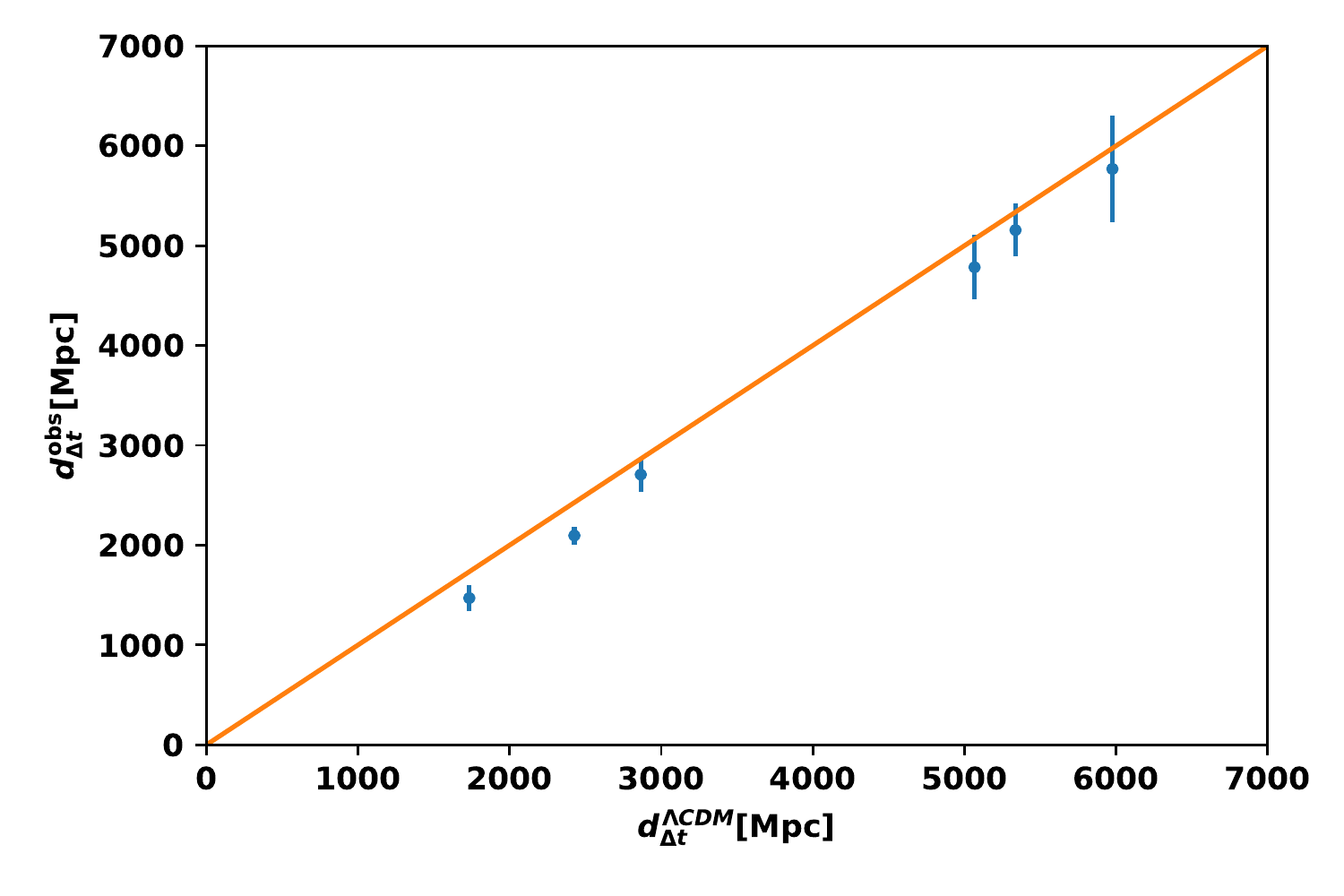,height=8cm,width=8cm,angle=0}
    \caption{A comparison of time-delay distance between observed values and a theoretical prediction based on flat $\Lambda$CDM model. }
    \label{fig:a5}
\end{figure}
Figure \ref{fig:a5} shows a diagonal test of time-delay distances between the observed H0LiCOW sample and the theoretical prediction by the flat $\Lambda$CDM model. From this figure, one can see that the values of time delay distances obtained from the flat $\Lambda$CDM model are higher than the  observed values at a given redshift.

\subsection{Further Comments}
$\bullet$ To study the expansion history of the universe at high redshifts, Gamma-Ray Bursts (GRBs) as standard candles are used beyond the existing reach of SN Ia observations. Nevertheless, the large dispersion in $E_{\mathrm{p}, \mathrm{i}}-E_{\mathrm{iso}}$ correlation,  limits the precision of distance determination with GRBs. Due to this reason, the use of GRBs as standard candles is highly controversial. Hence, in order to put strong constraints on cosmological parameters, we should look for more accurate luminosity relationships and investigate the classification problem of GRBs. 
\vspace{2mm}\\
$\bullet$ Our analysis indicates that the constraint on the cosmic curvature parameter is strongly dependent on the choice of the lens model of the galaxy. In the distance ratio method (with EPL lens model), the best fit value of the cosmic curvature parameter indicates a spatially closed universe and for time delay distance method (with SIS lens model), a spatially open universe is preferred. But a spatially flat universe is accommodated at $95\%$ confidence level in both methods. However, Wagner investigates that strong gravitational lensing observations' properties i.e. time delay differences, the relative image positions, relative shapes, and magnification ratios are invariant under the transformation of the source position and of the deflection angle\cite{jw2018}. They also identify that these transformations are confined to the regions of multiple images  as it helps to  constrain lens properties  without assuming a lens model. Due to the paucity of currently available data as well as the dependence of the lens parameters on the observations, the results of this work are dominated by the systematic errors of constraint parameters. However, we expect that ongoing and future surveys will provide more data on  SGL systems which will further help to improve the constraints on the cosmological parameters as well as lens profile parameters\cite{ag2020,pd2021}.
\vspace{1mm}\\
$\bullet$ As we mentioned, we assume $H_{0}=74.03 \pm 1.42 \mathrm{~km} \mathrm{~sec}^{-1} \mathrm{~Mpc}^{-1}$ throught the whole analysis and put constraints on lens parameters $(\gamma~\&~\delta)$ and cosmological parameters $(\Omega_{k0}~\&~\eta)$. To check dependency of cosmological parameters on $H_0$, we repeat the time-delay analysis with different values of Hubble constant: $H_0=67.66\pm0.42 \mathrm{~km} \mathrm{~sec}^{-1} \mathrm{~Mpc}^{-1}$\cite{planck2018} and $H_0=70.0\pm0.7 \mathrm{~km} \mathrm{~sec}^{-1} \mathrm{~Mpc}^{-1}$ [Fiducial Value]. The best fit values of $\Omega_{k0}$, $H_0$ and the model parameter $\eta_1$ for the two parametrisations are shown in Table \ref{tb.r1} and Table \ref{tb.r2}.
\begin{table}[ht]
\centering
\renewcommand{\arraystretch}{2}
     \begin{tabular}[b]{| c | c |c| c|}\hline
       Parameter &  $H_0=67.66\pm0.42$ & $H_0=70.0\pm0.7$\\ \hline \hline
    $\Omega_{k0}$ &  $0.332^{+0.192}_{-0.208}$ & $0.329^{+0.193}_{-0.204}$  \\ \hline
   $\eta_1$ & $0.281^{+0.209}_{-0.171}$   & $0.262^{+0.202}_{-0.158}$ \\ \hline
 
    \end{tabular}
\caption{ Results for $\eta_I=1+\eta_1z$. }
\label{tb.r1}
\end{table}
\vspace{0.2mm}
\begin{table}[ht]
\centering
\renewcommand{\arraystretch}{2}
     \begin{tabular}[b]{| c |c| c|}\hline
       Parameter &  $H_0=67.66\pm0.42$ & $H_0=70.0\pm0.7$\\ \hline \hline
    $\Omega_{k0}$ & $0.112^{+0.190}_{-0.143}$ & $0.108^{+0.191}_{-0.141}$  \\ \hline
   $\eta_1$ & $0.414^{+0.232}_{-0.221}$   & $0.396^{+0.212}_{-0.208}$ \\ \hline
    \end{tabular}
\caption{ Results for $\eta_{II}=1+\eta_1\dfrac{z}{1+z}$. }
\label{tb.r2}
\end{table}

Comparing the results tabulated in Table \ref{tb:a3} and Table \ref{tb:a4} to results shown in Table \ref{tb.r1} and Table \ref{tb.r2}, we find that for these three values of $H_0$, the values of $\Omega_{k0}$ obtained in the two models do not change substantially. They are all well within $68\%$ confidence level of each other. 
\vspace{2mm}\\
$\bullet$ Recently, based on the Broken Power Law (BPL) density profile, Du et al.  have developed an analytic model for the lensing mass of galaxies \cite{wd2019}. In their analysis, they claim that the high efficiency and accuracy of their model gives a promising method for analyzing galaxy properties with strong lensing. Therefore, by studying BPL density profiles of lens galaxies with observations, one may put a better constraint on the cosmic curvature parameter.

\section*{Acknowledgments}
We are grateful to the anonymous reviewer for her/his very enlightening remarks which have helped improve the paper. Darshan is supported by an INSPIRE Fellowship under the reference number: IF180293, DST India. NR and Darshan acknowledge facilities provided by the ICARD, University of Delhi. In this work some of the figures were created with \textbf{ {\texttt{corner}}} \cite{corner},  \textbf{ {\texttt{numpy}}}  \cite{numpy}  and \textbf{{\texttt{ matplotlib}}} \cite{matplotlib} Python software packages and to estimate parameters we used the publicly available MCMC algorithm  \textbf{ {\texttt{emcee}}} \cite{emcee}.

\end{document}